\documentclass[iop]{emulateapj}

\def\lapp{\ifmmode\stackrel{<}{_{\sim}}\else$\stackrel{<}{_{\sim}}$\fi}
\def\gapp{\ifmmode\stackrel{>}{_{\sim}}\else$\stackrel{>}{_{\sim}}$\fi}
\usepackage{multirow}
\usepackage{color}
\usepackage{amsmath}
\usepackage{soul}

\begin{document}

\title{Flux Relaxation after two outbursts of the magnetar SGR~1627$-$41 and possible hard X-ray emission}

\author{
Hongjun An\altaffilmark{1,*}, Andrew Cumming\altaffilmark{2}, and Victoria M. Kaspi\altaffilmark{2}
\\
{\small $^1$Department of Astronomy and Space Science, Chungbuk National University, Cheongju, 28644, Republic of Korea}\\
{\small $^2$Department of Physics \& McGill Space Institute, Rutherford Physics Building,
McGill University, 3600 University Street, Montreal, Quebec, H3A 2T8, Canada}\\
}
\altaffiliation{$^*$hjan@chungbuk.ac.kr}

\begin{abstract}
	We report on the long-term flux relaxation of the magnetar SGR~1627$-$41 after
its 2008 outburst, and evidence for hard X-ray excess measured with {\it NuSTAR}.
We use new observations made with {\it Chandra} and {\it XMM-Newton},
and an archival {\it NuSTAR} observation which add flux measurements
at $\sim$2000 days into quiescence after the 2008 outburst.
We find that the source flux has further declined since the last measurement
made in 2011, $\sim$1000\,days after the outburst in 2008. This trend is similar
to the relaxation after the source's 1998 outburst.
We use crustal cooling models to reproduce the flux relaxation;
if the whole surface of the star is heated
in the outbursts, the modeling suggests that the 2008 outburst
of SGR~1627$-$41 deposited energy into the inner crust and that
the core temperature of SGR~1627$-$41 is low ($T_c \lesssim 10^8$\,K) as previously
suggested. On the other hand, if only a small fraction
of the surface is heated or the temperature in the crust reached the melting temperature,
relaxation at early times requires another emission mechanism.
Finally, we report on evidence for hard X-ray emission in
SGR~1627$-$41 which follows the observational correlation suggested by \citet[][]{kb10}
in magnetars.
\end{abstract}


\section{Introduction}

	Magnetars are neutron stars which have very strong magnetic fields \citep{td95,td96},
typically $B>10^{14}$\,G, as inferred from their spin properties,
with some exceptions \citep[SGR~0418+5729, Swift~J1822.3$-$1606;][]{reti+10,skc14}.
Their emission is almost all in the X-ray band and is
believed to be produced by decay of the enormous internal magnetic fields. Hence,
magnetars often have larger X-ray luminosity than their rotational power.
They exhibit diverse observational properties \citep{ok14,mpm15,kb17} such as
short soft gamma-ray bursts, more dramatic long X-ray outbursts and giant flares,
and spectral turn-over in the hard X-ray band ($\sim$10\,keV).
See also the online magnetar catalog for various observational properties of
magnetars.\footnote{http://www.physics.mcgill.ca/∼pulsar/magnetar/main.html}

	Emission from magnetars is believed to be supported by the strong magnetic field;
internal decay of the magnetic field produces heat which is released at the surface as
thermal emission \citep[][]{tlk02}. The thermal spectrum is further modified in the
atmosphere due to absorption or in the magnetosphere due to resonant scattering \citep[][]{ft06},
and a non-thermal tail is seen at higher energies. Intriguingly, rising trends in their
spectral energy distributions (SEDs)
above 10\,keV were seen in some magnetars \citep[][]{khdc06}.
The possible origin of this emission was discussed by several authors \citep[][]{hh05,tb05,bh07,bt07,wbgh17},
and recently magnetar hard X-ray emission was investigated with
a coronal outflow model \citep[][]{bel13}
for some magnetars \citep[e.g., 1E~1841$-$045, 4U~0142+61, 1E~2259+586;][]{ahkb+13,vhka+14,thyk+15}, 
and was used for inferring the emission geometry.
Observationally, for hard X-ray bright magnetars, \citet{kb10} found a correlation between the
degree of spectral break and the spin-inferred magnetic field strength, which can give
further insights into the hard X-ray emission mechanism \citep[e.g.,][]{esks+17}.

	The mechanism by which the magnetic energy is released in magnetars is still
uncertain \citep[][]{bl16}. As the strong magnetic fields inside a magnetar evolve,
the external field can be twisted and the crust distorted, depositing energy both
inside and outside the star \citep{tlk02}. Energy deposited in shallow regions
of the crust may explain the observed surface luminosities of persistent
magnetars \citep[][]{kyps+06, bl16}, and may be responsible for the transient
emission seen in magnetar outbursts, in which the X-ray luminosity of the star
increases by orders of magnitude. These outbursts have been modeled as due
to release of stress in the crust by temperature-dependent plastic
motions \citep[][]{ll12,bl14,tyo16}. Magnetospheric activity can also lead to
energy being deposited in the outer layers of the star \citep[][]{lb15}.

\begin{table*}[t]
\vspace{-0.0in}
\begin{center}
\caption{Summary of observations used in this work
\label{ta:ta0}}
\vspace{-0.05in}
\scriptsize{
\begin{tabular}{ccccccc} \hline\hline
Observatory      & ObsId  & Start time &  Instruments    & observation mode  &   Exposure    &  Comment   \\
                 &        &  (MJD) &                 &                  &        (ks)    &   \\ \hline
{\it Chandra}    &  15625 	 & 56374  &    ACIS-I    &  Imaging       &   10   & 10$'$ off-axis \\ 
{\it XMM-Newton} &  0742650101	 & 57071 & MOS1,MOS2,PN & Full window  &   40   & medium filters \\ 
{\it NuSTAR}     & 30160002002	 & 57181 &  FPMA, FPMB  & $\cdots$  &  100   &  $\cdots$   \\ \hline
\end{tabular}}
\end{center}
\vspace{-0.5 mm}
\end{table*}

	The transient relaxation of a magnetar after the outburst can give clues
as to the nature and location of the heating that causes the outburst,
as well as the response of the star and the magnetosphere.
The flux relaxation after outburst has been modeled by both crust cooling
and untwisting of the
external magnetosphere. In the crust cooling models \citep[e.g.,][]{let02,kewl+03,pr12,skc14},
energy is deposited inside the star, and the subsequent cooling is calculated assuming
crust properties and the energy deposition profile.
In untwisting models \citep[e.g.,][]{b09}, the outburst suddenly twists
the external magnetic field in a specified way, and then the flux relaxes
to the quiescent level as the twist unwinds. Both mechanisms could contribute
to the transient relaxation, but with different trends over time.

	SGR~1627$-$41 (hereafter SGR1627) was discovered during
an outburst in 1998 \citep{wkvh+99} and exhibited another outburst in 2008 \citep{eizs+08}.
The spin-down rate ($P=2.59$\,s) and its first
derivative ($\dot P=1.9\times10^{-11}$) were measured after the 2008 outburst
when the source was bright \citep{ebpt+09}. This implies that the surface dipolar magnetic field
strength is $B=2\times10^{14}$\,G, well within the range for magnetars.
The flux relaxations after the outbursts have been relatively well
measured \citep{kewl+03,eizs+08,aktc+12}. 

	The fact that two outbursts have been observed from the same source
is interesting for the crust cooling model because it could remove some of the
degeneracy between the parameters of the model; the shape of the cooling curve depends
on both crust properties and the energy deposited, but the crust properties should be
the same from outburst to the next. In addition, the long timescales of thousands of
days over which the flux decay has been observed make the light curves possibly sensitive to the
interesting physics of the crust/core boundary. For example, it has been suggested that
the pasta phase expected at these densities may give a low
conductivity \citep[][]{pvr13,hbbc+15}. Because the published observations of the decay
following the most recent 2008 outburst extend to $\sim 1000$ days, with the possibility
that the source could further drop in flux, the interpretation in the context of the crust
cooling model is uncertain. With different assumptions about the heating profile,
\cite{aktc+12} concluded that the neutron star core temperature could be higher than
previously thought, whereas \cite{dcbr17} assumed a lower core temperature and
made predictions for the future flux evolution that depended on the parameters chosen
for the inner crust.

	We report on new observations of SGR1627 made with {\it Chandra} and {\it XMM-Newton}
and an archival {\it NuSTAR} observation.
These observations sample very late-time transient relaxation of the source,
$\gapp 2000$ days into quiescence, for which modeling
the relaxation trend provides a new test
of the crust cooling and the magnetospheric untwisting models,
and may give us new insights into the
energy deposition in outbursts of magnetars.
In section~\ref{sec:sec2}, we describe the observational data and
present our analysis results. We show our modeling results in
section~\ref{sec:sec3} and then discuss and conclude in section~\ref{sec:sec4}.

\section{Data Analysis and Results}
\label{sec:sec2}

\begin{figure*}
\hspace{0.0 mm}
\begin{tabular}{cc}
\includegraphics[width=3.4 in,angle=0]{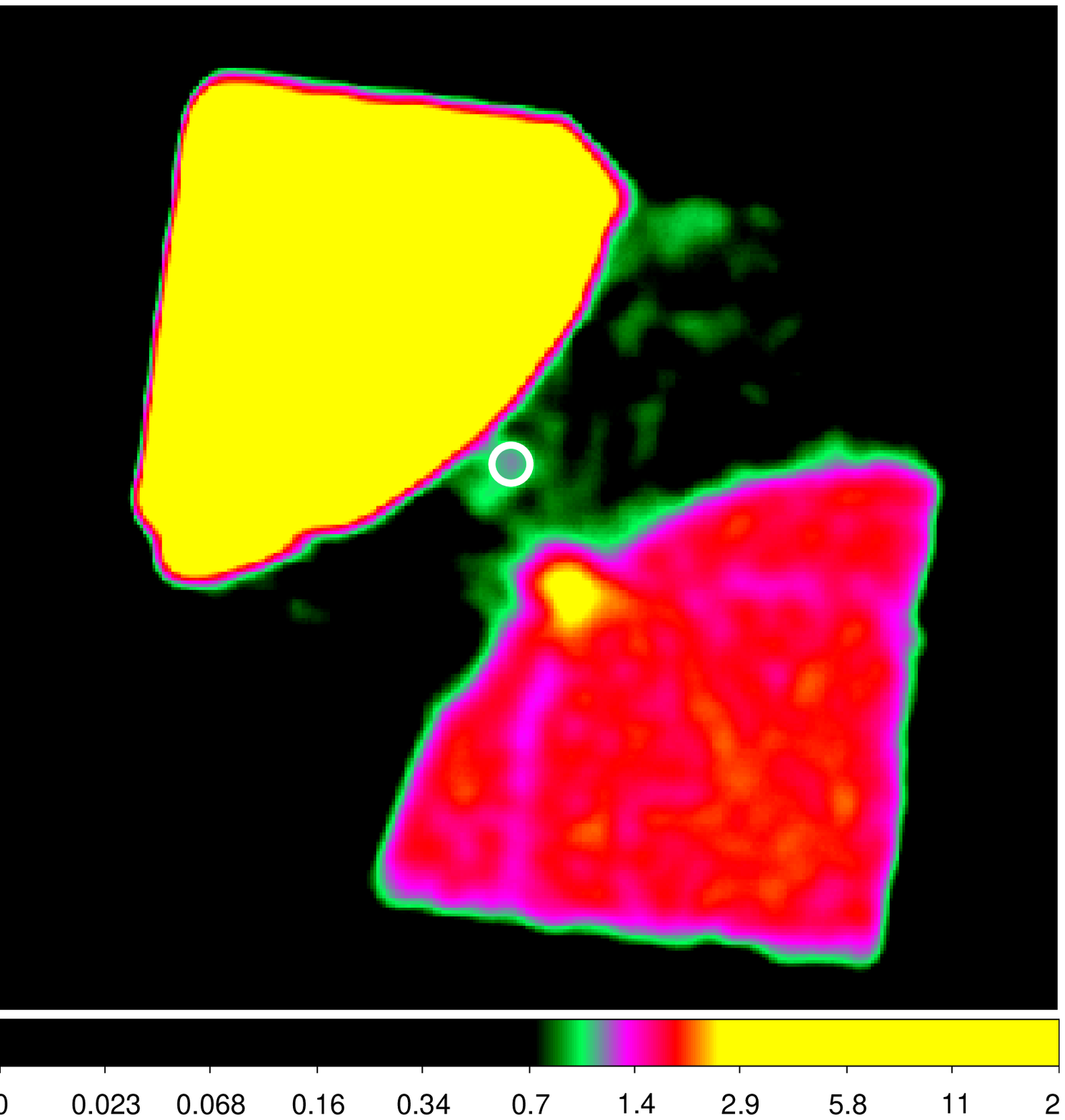} &
\includegraphics[width=3.4 in,angle=0]{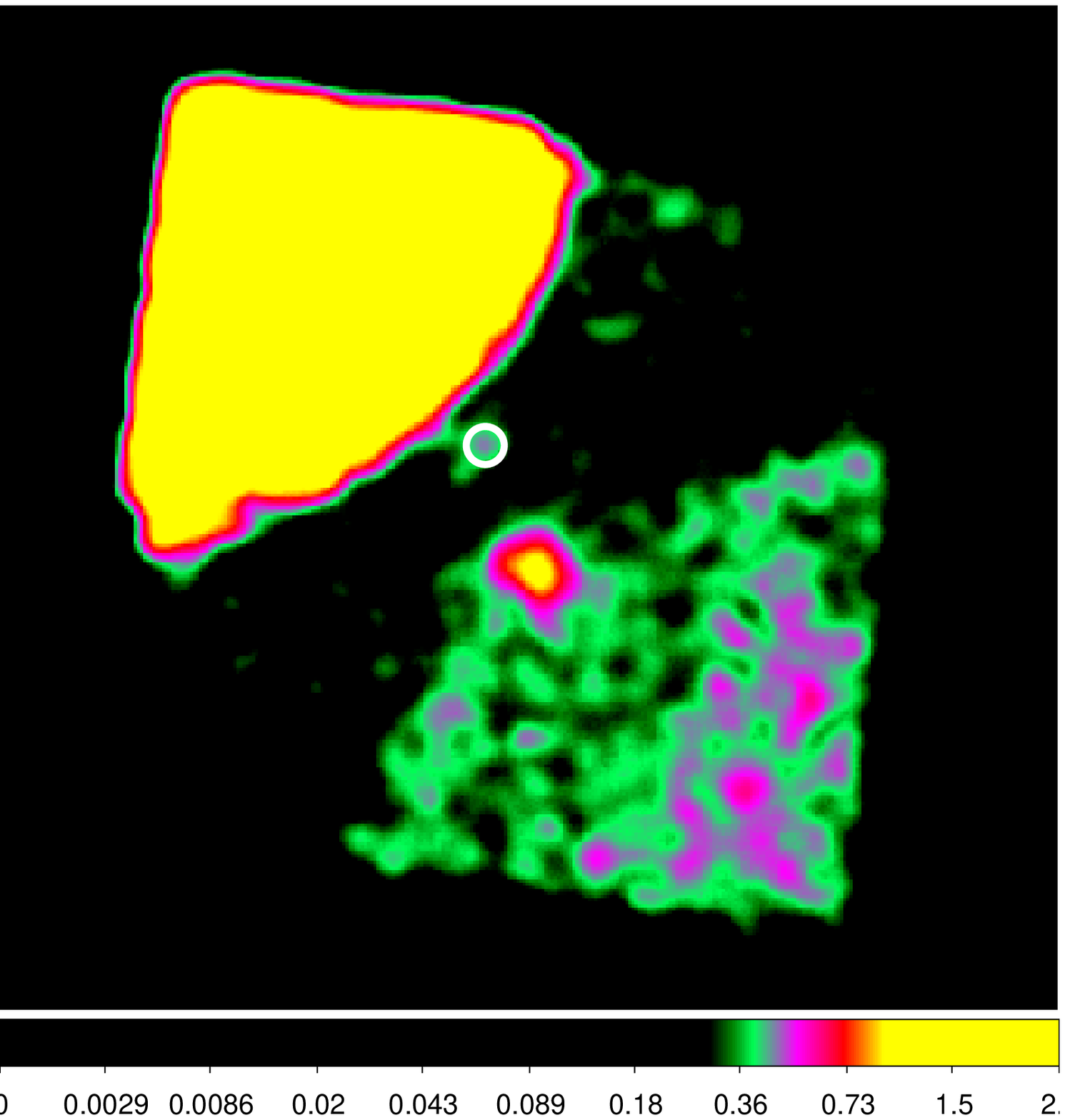} \\
\vspace{0.0 mm}
\end{tabular}
\figcaption{Soft- and hard-band images of the {\it NuSTAR} observation in the 3--10\,keV (left)
and 10--20\,keV bands (right). The magnetar SGR1627 is shown in a circle.
The image is smoothed and the color scale is changed for better legibility.
The bright point source $\sim$2$'$ south-west of SGR1627 is CXOU~J163547.0$-$473739
\citep[][]{fthg+17}, and the large bright patterns in the north east and south west
are produced by stray light (photons that are not reflected by the focusing mirrors)
or ghost rays (photons that are reflected only once)
from the nearby (but outside the FoV) bright sources GX~340+0 and 4U~1624$-$49 \citep[][]{fthg+17}.
See \citet{mccf+17} for more details about stray light and ghost rays.
\label{fig:fig1}
}
\vspace{0mm}
\end{figure*}

	We used a 10-ks {\it Chandra} and a 40-ks {\it XMM-Newton} observation made on
MJDs~56374 and 57071, respectively (Table~\ref{ta:ta0}). The {\it Chandra} data are reprocessed
with {\tt chandra\_repro} of CIAO~4.8 along with CALDB~4.7.2 to use the
most recent calibration files, and the 40-ks {\it XMM-Newton} data are processed
with {\tt epproc} and {\tt emproc} of SAS~20160201\_1833-15.0.0.
Note that in the {\it Chandra} observation SGR1627 was observed with a $\gapp$10$'$
offset from the optical axis, hence the effective photon-collecting area was small,
and only a few events were detected near the source position.
We also analyzed archival 100-ks {\it NuSTAR} data \citep[][]{fthg+17},
taken $\sim$100\,days after the {\it XMM-Newton} observation (Figure~\ref{fig:fig1}).
The data were processed with {\tt nupipeline} integrated in the {\tt HEASOFT}~6.19 along
with the HEASARC remote CALDB. We further process the data for analyses below.
We also use published results \citep{kewl+03,eizs+08,aktc+12}
for constructing and modeling SGR1627's flux relaxation trends.

\subsection{{\it XMM-Newton} Data Analysis}
\label{sec:sec2_1}
	We measure the spectrum of SGR1627 with the {\it XMM-Newton} data because
these provide by far the best statistics. Although SGR1627 was in its late stage of
transient cooling and thus is very faint, it is visible in each of MOS and PN
exposures. We extracted source events within a $R=16''$ circle around
the source position, collecting 310 events (PN and MOS1,2 combined in the 0.5--10\,keV band) in this region.
Background events were extracted in a source-free $R=32''$ circular region, 200$''$ north
of SGR1627 (580 events); the detection significance is $\sim$8$\sigma$.
We calculated spectral response files using the {\tt rmfgen}
and the {\tt arfgen} tools of SAS.
We group the spectra to have at least 5 counts per energy bin
and use an absorbed power-law or an absorbed blackbody model. Unless otherwise stated,
we use $\tt wilm$ abundance \citep[][]{wam00} and $\tt verner$ cross
section \citep[][]{vfky96} because these are used for cross calibration of the
{\it NuSTAR} data with other X-ray satellites \citep[][]{mhma+15}.

	We fit the spectra using {\it l}-statistic in {\tt XSPEC}~12.9.0n \citep{l92}.
The best-fit $N_{\rm H}$ and $\Gamma$ are consistent with the previous results \citep[e.g.,][]{eizs+08}.
We, however, find that the source flux declined by a factor of 2--3 compared with the last
measurement of $F_{\rm 2-10\ keV}=2.7\pm0.4\times10^{-13}\rm\ erg\ cm^{-2}\ s^{-1}$ made in 2011.
The significance of the flux drop is $\sim$3$\sigma$ if we fit $N_{\rm H}$ as well, and larger
if we hold $N_{\rm H}$ fixed at the value measured by \citet{eizs+08}.
The blackbody model can also explain the data, with high surface temperature compared with other
magnetars in quiescence. Because the power-law model is slightly better and the previous measurements
were all done with that model, we use the power-law flux for the light curve below.
The results of the fitting are summarized in Table~\ref{ta:ta1}.
We also used a different absorption model, {\tt angr} abundance and the {\tt bcmc}
cross section \citep[][]{angr89,bcmc92} for comparison with the 2011 measurements,
and find that the best-fit parameters do not
change significantly. In this fit, the best-fit $N_{\rm H}$ is smaller and the 2--10\,keV
flux is $\sim$10\% larger.

\newcommand{\marka}{\tablenotemark{a}}
\newcommand{\markb}{\tablenotemark{b}}
\newcommand{\markc}{\tablenotemark{c}}
\begin{table*}[t]
\vspace{-0.0in}
\begin{center}
\caption{Fit results for the {\it XMM-Newton} and the {\it NuSTAR} data
\label{ta:ta1}}
\vspace{-0.05in}
\scriptsize{
\begin{tabular}{ccccccccc} \hline\hline
Model\marka & Inst. & Energy  & $N_{\rm H}$              &   $\Gamma/kT$     & $\Gamma_2$ & $E_B$ & Flux         & $lstat$/dof       \\
            &       & (keV)   &   ($10^{22}\rm cm^{-2}$) &  ($\cdots$/keV)    &  &  (keV)  &               &                \\ \hline
PL          & X     & 0.3--10 &  $13\pm5$                &  $2.6\pm0.8$ & $\cdots$  & $\cdots$  & $10^{+4}_{-3}$      & 41.6/54   \\
PL          & X     & 0.3--10 &  $10\markc$              &  $2.0\pm0.3$ & $\cdots$  & $\cdots$  & $8\pm1$             & 42.4/55   \\ \hline
BB          & X     & 0.3--10 &  $7\pm3$                 &  $1.3\pm0.2$ & $\cdots$  & $\cdots$  & $6\pm1$             & 43.5/54   \\
BB          & X     & 0.3--10 &  $10\markc$              &  $1.1\pm0.1$ & $\cdots$  & $\cdots$  & $7\pm1$             & 44.6/55   \\ \hline
PL          & N     & 3--20 &    $10\markc$              &  $1.0\pm0.6$ & $\cdots$  & $\cdots$  & $4^{+2}_{-1}$       & 53.2/55   \\ 
PL          & N     & 3--10 &    $10\markc$              &  $3.4\pm1.0$ & $\cdots$  & $\cdots$  & $6^{+3}_{-2}$       & 33.1/39   \\ \hline
PL          & X+N   & 0.3--20 &  $10\markc$              &  $1.8\pm0.3$  & $\cdots$  & $\cdots$  & $6^{+2}_{-1}$      & 97.7/112   \\ 
BPL         & X+N   & 0.3--20 & $10\markc$               &  $2.2\pm0.3$ & $-0.4\pm1.2$  & $8.9\pm2.6$  & $4^{+2}_{-1}$ & 88.8/110   \\ \hline
\end{tabular}}
\end{center}
\vspace{-0.5 mm}
$^{\rm a}${PL: power law, BB: Blackbody.}\\
$^{\rm b}${Absorption corrected flux in units of $10^{-14}\rm erg\ cm^{-2}\ s^{-1}$ in the
2--10\,keV band for {\it XMM-Newton} and in the 3--10\,keV band for {\it NuSTAR}
or {\it XMM-Newton} + {\it NuSTAR}.}\\
$^{\rm c}${Frozen.}\\
\end{table*}

\subsection{{\it Chandra} Data Analysis}
\label{sec:sec2_2}
	As we noted above, there are only a few
events detected near the source position in the {\it Chandra} data, and hence we measure
a flux upper limit only. For this, we used a circular aperture with $R=5''$ for
the source and an annular aperture with $R_{\rm in}=10''$ and $R_{\rm out}=20''$ for
the background. In these apertures, we find 1 event in the source region and
16 events in the background region in the 2--10\,keV band.
Since it is impractical to perform a spectral fit with
these small numbers of events, we measure a flux upper limit.

\begin{figure*}
\hspace{3.0 mm}
\includegraphics[width=6.5 in,angle=0]{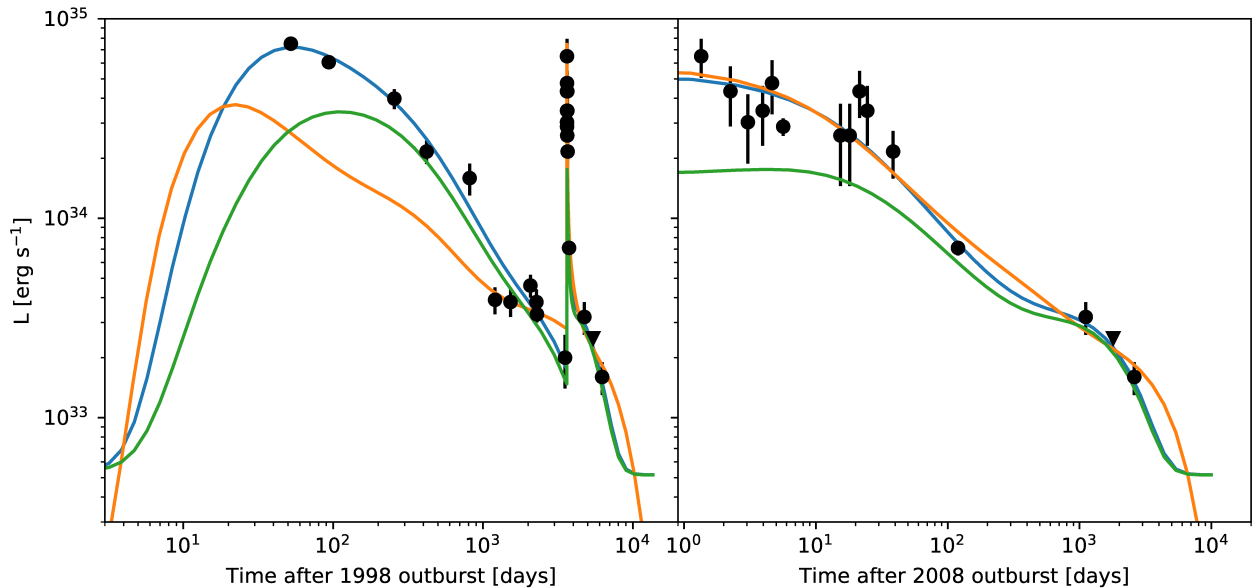}
\vspace{0.0 mm}
\figcaption{Flux relaxation data after the 1998 (left) and the 2008 (right) outbursts and 
crust cooling models (solid lines) for SGR1627. {\em Blue}: heating of the whole
surface as in \citet{dcbr17}. To match the 2008 outburst at late times, we include heating
in the inner crust. {\em Orange}: only 10\% of the neutron star surface is heated,
resulting in larger crust temperatures. The early part of the light curve reaches
a maximum luminosity set by neutrino emission \citep[][]{pr12}. By reducing the
neutron star surface gravity, we are able to fit the late time cooling without
any inner crust heating in the 2008 outburst in this case.
{\em Green}: Same as blue, but with the crust temperature limited to be less
than the melting temperature. The luminosities are calculated with 2--10\,keV fluxes
assuming a distance of 11\,kpc \citep[][]{ccdd99}.
The details of the heating profile for each model are given in the caption of
Figure~\ref{fig:fig4}.
\label{fig:fig2}
}
\vspace{0mm}
\end{figure*}

	We first calculated
the effective area for the source region using the {\tt specextract} tool of CIAO and used the
{\it XMM-Newton}-measured power-law spectral shape as our source model because a power-law model
fits the {\it XMM-Newton} data better than a blackbody model does,
and the previous results are made with a power-law model.
We then adjust the normalization of the source model (i.e., flux) and fold it
with the effective area until the
chance probability (Poisson statistic) of having 1 or less event within a $R=5''$ circle
in the 2--10\,keV band is smaller than 10\%, considering the background as well.
We find this flux value to be $1.7\times 10^{-13}\rm \ erg\ s^{-1}\ cm^{-2}$ and set
this as the 90\% upper limit for the source flux at the observation epoch. Note that
this is also smaller than the 2011 measurement. Our measurements of the source fluxes
are shown in Figure~\ref{fig:fig2}. 

\subsection{{\it NuSTAR} Data Analysis}
\label{sec:sec2_3}
	In the {\it NuSTAR} observation, SGR1627 was very faint,
dominated by background.
The magnetar was detected only at the $\lapp$6$\sigma$ level in the 3--25\,keV band, and so
the data are insufficient for determining the spectrum accurately.
However, it can provide a consistency check for the {\it XMM-Newton} results, and thus
we perform a spectral analysis in the 3--10\,keV band.
We extracted spectra using $R=20''$ apertures for two {\it NuSTAR} modules FPMA and FPMB,
and generated response files using the {\tt nuproducts} tool.
We then bin the spectra to have at least 5 events per
spectral bin and fit the spectra in {\tt XSPEC} with an absorbed power-law model using
the $l$ statistic.

	The power-law index we measure is somewhat smaller (but not significantly)
than that reported by \citet{fthg+17}, but agrees with the {\it XMM-Newton} result.
The measured spectral index and the 2--10-keV-extrapolated flux are $3.4\pm1.0$ and
$12\times 10^{-14}\rm \ erg\ cm^{-2}\ s^{-1}$ (Table~\ref{ta:ta1}).
We did not attempt to fit the {\it NuSTAR} data with a blackbody model because of the
lack of sensitivity of {\it NuSTAR} below 3\,keV.
Because the background dominates the spectra in this analysis,
the results may be sensitive to the background
selection. We therefore tried with backgrounds taken from different regions
and found that the results do not
change significantly. We report the best-fit parameters in Table~\ref{ta:ta1}.
Note that the uncertainty for the flux measurement is large, so we check to see if
it is consistent with zero by scanning the photon index and flux with the {\tt steppar} command
of {\tt XSPEC}. We find that the flux is not consistent with zero at the 99\% confidence level.

\begin{figure}
\hspace{-3.0 mm}
\includegraphics[width=3.3 in]{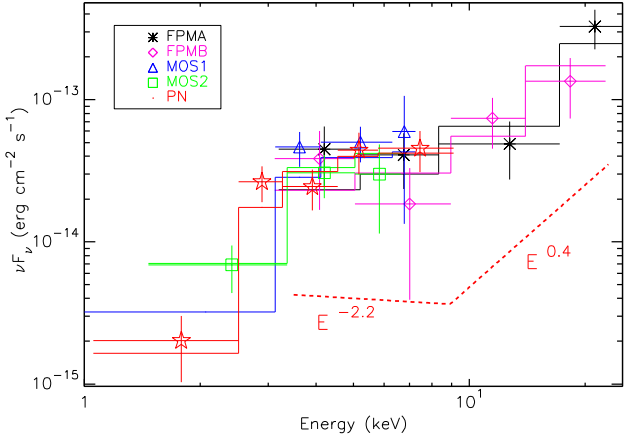}
\figcaption{Broadband X-ray spectral energy distribution (SED, $E^2\frac{dN}{dE}$)
measured with {\it XMM-Newton} and {\it NuSTAR}
(points with error bars) and the best-fit broken power-law model (solid lines).
The spectra measured with {\it XMM-Newton} are shown
in blue (MOS1), green (MOS2) and red (PN). {\it NuSTAR} spectra are shown in
black (FPMA) and magenta (FPMB). We also show a broken power-law line (red dashed)
separately for clarity.
\label{fig:fig3}}
\vspace{3 mm}
\end{figure}

\subsection{Broadband X-ray Spectral Analysis}
\label{sec:sec2_4}
	In the {\it NuSTAR} data analysis, we find evidence for spectral hardening with energy,
as has been seen in some magnetars \citep[][]{khdc06,vhka+14,aahk+15,thyk+15}. The spectral
index for a simple power-law model is $1\pm0.6$ when fitting the data in the 3--20\,keV band,
much smaller than $3.4\pm1.0$ measured in the soft 3--10\,keV band (Table~\ref{ta:ta1}).
This trend can be seen in the spectrum (Figure~\ref{fig:fig3}). We note that the
source is clearly detected up to $\sim$20\,keV \citep[see also][]{fthg+17},
and is brighter compared to the background at higher energies than at lower energies.
In order to investigate this further, we fit the {\it NuSTAR} and the {\it XMM-Newton}
data jointly. Although the observations are separated by $\sim$100\,days,
no large outburst has been seen during this time and the soft-band source
spectrum seems not to have changed; the 3--10\,keV {\it NuSTAR} spectra are consistent
with the {\it XMM-Newton} one; so jointly fitting the spectra is unlikely to suffer
from systematic errors due to different flux levels.

	We fit the 0.5--10\,keV {\it XMM-Newton} and the 3--20\,keV {\it NuSTAR} data jointly
with an absorbed power-law or an absorbed broken power-law model, holding $N_{\rm H}$ fixed
at $10^{23}\rm \ cm^{-2}$. Due to the paucity
of counts, both models explain the data reasonably when using the $\chi^2$ statistic with {\tt gehrels}
weighting \citep[][]{gehrels86}; $\chi^2$/dof's are 45.17/112 and 41.55/110 for the power-law and
the broken power-law models, respectively, and the $f$-test probability that the additional
parameters are unnecessary is 0.01. Because we are concerned with the results being sensitive
to the statistics employed, 
we further test the fits with the $l$ statistic (5 cts per bin) or
$c$ statistic \citep[1 cts per bin;][]{cash79}. For these, we find that the
best-fit spectral parameters are consistent with those obtained with the $\chi^2$ fit (Table~\ref{ta:ta1}),
and the Akaike information criterion \citep[][]{aic74} also tells us
that the broken power-law model is favored over the simple power-law model.

\subsection{Timing Analysis}
\label{sec:sec2_5}
	We perform a timing analysis to search for pulsations with the
{\it XMM-Newton}/PN and the {\it NuSTAR} data which were both taken with sufficient
timing resolution (72\,ms for {\it XMM-Newton}/PN and $\sim$2\,ms for {\it NuSTAR}).
The pulse period and the first derivative of SGR1627 were
measured to be $P=2.59439$\,s and $\dot P=1.9\times 10^{-11}$ on MJD~51259 \citep{ebpt+09}.
When extrapolated to the epochs of the observations we are using,
the expected period is $P\simeq2.60$\,s. We therefore searched for pulsations between $P=2.5$\,s
and $P=2.7$\,s using the $H$ test \citep{drs89}, but found no significant pulsation in the data.
This is not surprising given the fact that the pulsations were detected only when the source
was bright after the 2008 outburst; the pulsed fraction is low \citep[0.13;][]{ebpt+09}
and so the signal-to-noise ratio is not likely high enough for a detection when the source is faint.
	
\section{Modeling the Flux Relaxation of SGR1627}
\label{sec:sec3}

	We compile data taken for two outbursts of SGR1627 to study the source's transient relaxation.
We added the measurements we made in this work to the data reported by \citet{kewl+03},
\citet{eizs+08} and \citet{aktc+12}.
The new observations sample the 2008 cooling trend at $\sim$1800\,days
and $\sim$2500\,days, and $\sim$2600\,days, (Figure~\ref{fig:fig2}) providing constraints
on the late time cooling, which depends on the properties of the inner crust
and magnetosphere of SGR1627,
and energy deposition profiles of the outbursts, and/or the magnetospheric twist/untwist.
We first summarize previous work on crust cooling applied to SGR1627 and then present
new models that fit the latest flux measurements.

\subsection{Previous work}
\label{sec:sec3_1}
	\citet{kewl+03} were able to reproduce the flux decay of SGR1627 following
its outburst in 1998 using a crust cooling model in which they deposited
$\sim 10^{44}\ {\rm ergs}$ in the crust. The temperature profile they assumed right after
the outburst was guided by the variation of specific heat capacity with depth
(in the outer crust, the specific heat is dominated by the nuclear lattice,
whereas it is strongly suppressed in the inner crust when the neutrons become superfluid).
They chose a low core temperature $\ll 10^8\ {\rm K}$ to reproduce the drop in luminosity
seen at late times (see their Fig.~1).

	After the second outburst in 2008, \citet{aktc+12} compared the 2--10\,keV
luminosity decays of both outbursts with crust cooling models. They considered models
in which energy is deposited instantaneously in the outer crust down to a density
$\rho_{\rm max}$, and then the crust is allowed to thermally relax.
They assumed spherical symmetry so that the heating is over the whole stellar surface, and neutron star parameters
$M=1.3\ M_\odot$, $R=12\ {\rm km}$, $B=2\times 10^{14}\ {\rm G}$, $Q_{\rm imp}=3$,
where $M$ and $R$ are the neutron star mass and radius, $B$ is the polar magnetic field,
and $Q_{\rm imp}$ is the impurity parameter that determines the conductivity
of the inner crust \citep[][]{ik93}. Modeling the 1998 and 2008 outbursts,
they found that the amount of energy deposited
must be about ten times greater
(energy density $E_{25} = E/10^{25}\ {\rm erg\ cm^{-3}}\sim 1$ for 2008
and $E_{25}\sim 10$ for 1998), and deeper
(up to a density $\rho_{\rm max}\approx 3\times 10^{10}\ {\rm g\ cm^{-3}}$ for 2008
and $\rho_{\rm max}\approx 2\times 10^{11}\ {\rm g\ cm^{-3}}$ for 1998)
for the 1998 outburst compared to the 2008 outburst, to match its brighter
and longer outburst. With these choices the shape of the outbursts could
be reproduced for times $\lesssim 1000$ days.

	\citet{aktc+12} noted that the difference in outburst decay times has implications
for the long timescale behavior. The luminosity at times $\approx 1000$ days was similar
for both outbursts. But the 2008 outburst decay was rapid enough and the heating
shallow enough that the crust should have come back into thermal equilibrium
with the core at $\sim1000$\,days. This implies a hot core with core temperature
$T_c\approx 2\times 10^8\ {\rm K}$ \citep[][]{aktc+12}. However, the 1998 outburst
showed a further drop in luminosity by a factor of two that was observed in an observation
taken several years after the outburst. This measurement is unexpected for a hot core,
which should keep the crust temperature high, but the observed luminosity is within
about 2-$\sigma$ of the previous value.

	\citet{dcbr17} investigated whether different choices for the (poorly constrained)
physics of the inner crust could explain the drop in luminosity in the last measurement
after the 1998 outburst. In particular, they looked at the effect of a low thermal
conductivity in the part of the inner crust that is thought to consist of a nuclear
pasta phase. A low electrical conductivity for the pasta phase had been suggested
by \citet{pvr13} in the context of neutron star magnetic field evolution, and the effect
of a corresponding low thermal conductivity in accreting neutron stars on thermal relaxation after accretion
outbursts had been calculated by \citet{hbbc+15}. However, \citet{dcbr17} showed
that the pasta layer does not delay cooling for the rapid heating thought to occur
in a magnetar outburst (as opposed to the heating by accretion that occurs on
timescales of years).

	Instead, \citet{dcbr17} found that the observations of \citet{aktc+12}
could be made consistent with a colder core if the inner crust was also heated in
the 2008 outburst. The amount of heating needed in the inner crust is comparable
to that in the 1998 outburst (since both outbursts reach similar luminosities
at times $\sim 1000$\,days). This means that the energetics of the two
outbursts are in fact similar, dominated by heat released in the inner crust,
but they have very different decay times because of the different energies deposited in the
outer crust. This picture is therefore quite different from the case with outer
crust heating only as in \citet{aktc+12}, where the two outbursts have
different energies by an order of magnitude.

\subsection{Models of the light curve}
\label{sec:sec3_2}

\begin{figure}
\hspace{-3.0 mm}
\vspace{0 mm}
\includegraphics[width=3.3 in]{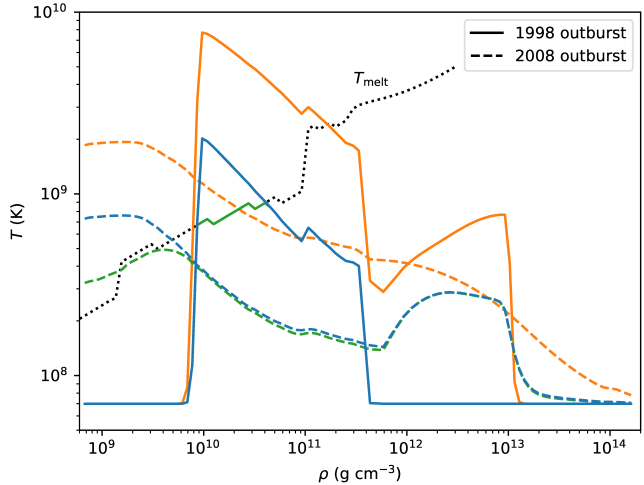}
\figcaption{Temperature profiles at the end of each outburst (1998 solid curves, 2008 dashed curves)
for the three models shown in Figure~\ref{fig:fig2}.
The blue curves have $M=1.4\ M_\odot$, $R=12\ {\rm km}$, and energy deposited over the entire stellar
surface with $E_{25}=E/10^{25}\ {\rm erg\ cm^{-3}}=13 (0.9)$
in the outer (inner) crust for the 1998 outburst and $E_{25}=0.9 (10)$
in the outer (inner) crust for the 2008 outburst
(total energies $6.1\times 10^{43}$ and $4.1\times 10^{43}\ {\rm ergs}$, respectively).
The green curves correspond to the same model as the blue curves, but with the temperature limited
to be less than the melting temperature (shown as the dotted curve in the Figure).
The orange curves have $M=1.2\ M_\odot$, $R=12.5\ {\rm km}$, and energy deposited over
10\% of the neutron star surface at the magnetic pole, with $E_{25}=130 (130)$ in the
outer (inner) crust for the 1998 outburst and $E_{25}=4.0 (0.0)$ in the outer (inner) crust
for the 2008 outburst (total energies $1.6\times 10^{44}$ and $3.6\times 10^{42}\ {\rm ergs}$, respectively).
The rapid increase in $T_{\rm melt}$ just above $10^{11}\ {\rm g\ cm^{-3}}$
corresponds to the transition from $Z=28$ to $Z=44$ in the equilibrium crust model.
\label{fig:fig4}}
\vspace{1 mm}
\end{figure}

	Now that we have two new observations showing that the flux has decreased
further, we return to crust cooling models to see whether we can reproduce the
observed flux decays and constrain the required heating and
crust parameters\footnote{The code used to calculate the crust thermal relaxation
is available at \url{https://github.com/andrewcumming/crustcool}.}.
We model the two outbursts in the same calculation, i.e.~we take into account
the fact that the inner crust had not completely relaxed after the 1998 outburst
when the 2008 outburst occurred. The basic parameters are similar to those in \citet{aktc+12}:
$M=1.4\ M_\odot$, $R=12\ {\rm km}$, $B=2\times 10^{14}\ {\rm G}$, $Q_{\rm imp}=1$,
and core temperature $T_c=7\times 10^7\ {\rm K}$. It is important to note that when
comparing to data, we compare only to the $2-10\ {\rm keV}$ luminosity and ignore
the bolometric correction. Therefore our conclusions about energetics and whether
we can match the observed luminosities should be taken as lower limits.
We also do not consider or model the observed X-ray spectrum;
we instead focus on the luminosity and see whether crust cooling models are able
to reproduce the shape of the observed decay.

	We briefly discuss the physics input of our model.
The numerical grid extends from a density $\rho\approx 6\times 10^8\ {\rm g\ cm^{-3}}$
to the crust core boundary at $\rho\approx 1.4\times 10^{14}\ {\rm g\  cm^{-3}}$
(baryon density $n_b=0.08\ {\rm fm^{-3}}$). General relativistic corrections are
included as an overall redshift of the time or luminosity.
The composition of the crust is taken from the cold-catalyzed matter calculations
of \citet{hp94} and \citet{dh01}. We take the melting
point to be when $\Gamma=Z^2e^2/ak_BT=175$ \citep[][]{pc00},
where $Z$ is the nuclear charge and $a$ the inter-ion spacing.
We do not include the effects of the magnetic field on the electron equation of state
(this is important only for $\rho < 2\times 10^8\ {\rm g\ cm^{-3}} B_{15}^{3/2} (0.5/Y_e)$,
which lies off our grid). The heat capacity has contributions from electrons, the ion lattice,
and neutrons, with the neutron superfluid gap taken from \citet{sfb03}.
Neutrino emission from plasmon decay and pair annihilation \citep[][]{sswm+87},
neutrino bremsstrahlung \citep[][]{hky96,kppt+99},
and neutrino synchrotron \citep[][]{bhky97} is included.
We calculate the thermal conductivity tensor, including the effect of the quantizing
magnetic field, using the results for electron-phonon, electron-impurity, and electron-ion scattering summarized in
\citet{ppp15}.\footnote{Using the Fortran routine {\tt condegin13.f} from the website of A.~Potekhin, http://www.ioffe.ru/astro/conduct/}
For computational convenience, we use the envelope model results from \citet{py01}
to calculate the flux at the outer boundary
(although as discussed by \citet{skc14} this can lead to a tens of percent underestimate of the flux
because the outer layers of the grid are not necessarily isothermal at high temperatures).
In models which heat the entire surface of the neutron star, we solve an angle-averaged
heat equation, averaged over latitude assuming a dipole magnetic field, following
\citet{gh83}. We have checked that this matches adequately with the
more computationally intensive approach of following each local patch on the star
separately and then summing the light curves.

	We first assume spherical symmetry, i.e.~heating across the entire neutron
star surface, as assumed by \citet{aktc+12} and \citet{dcbr17}. The blue curve
in Figure~\ref{fig:fig2} shows a model similar to \citet{dcbr17}.
It has an order of magnitude difference in heating in the outer crust
($\rho<4\times 10^{11}\ {\rm g\ cm^{-3}}$) between the two outbursts:
$E_{25}=E/10^{25}\ {\rm erg\ cm^{-3}}=13$ for 1998 and
$E_{25}=0.9$ for 2008. To match the observed luminosity at times
$\gtrsim 1000$ days, we include a heating $E_{25}=10$ in the inner crust for the 2008 outburst.
No inner crust heating is included for the 1998 outburst, but it changes
the 1998 light curve only slightly. After the inner crust begins to cool,
the luminosity drops to a level set by the core temperature, here taken
to be $T_c=7\times 10^7\ {\rm K}$. We show the temperature profiles in the crust
after the outbursts in Figure~\ref{fig:fig4}.

	Although it is possible that the heating occurs over most of the surface of
the star \citep{tyo16}, the heated region could be substantially smaller.
The orange curve in Figure~\ref{fig:fig2} shows a model in which we heat only $10$\%
of the neutron star surface (the magnetic field is assumed
to be radial, so that the heated region is near the magnetic polar cap).
Because the heated area is ten times smaller, the local energy density deposited
in the crust is approximately an order of magnitude larger than in the previous model,
making the crust significantly hotter. This has two effects.
First, in the outer crust neutrino emission becomes important and limits the temperature.
This gives a maximum luminosity that can be achieved at times of tens to hundreds of days,
as discussed by \citet{pr12}. As can be seen in Figure~\ref{fig:fig2}, for this reason
we are not able to match the observed light curve at these early times.
The second effect of the high crust temperature is to increase the cooling time.
However, this is offset by the enhanced conductivity associated with the radial field lines
near the polar cap. We find that we can then match the decreasing luminosity at
$\approx 3000$ days by decreasing the neutron star gravity
(the model shown has $M=1.2\ M_\odot$ and $R=12.5\ {\rm km}$).
Inner crust heating is not needed for the 2008 outburst; however it is needed
in the 1998 outburst.

	Another possible limitation at high temperature is that the crust melts.
Whether this acts to limit the temperature depends on the mechanism of energy release.
If the outburst results from dissipation due to a thermoplastic wave, as discussed
by \citet{bl14} for example, the maximum temperature expected at a given depth is
the local melting temperature $T_{\rm melt}$. The green curve in Figure~\ref{fig:fig2}
shows a model with parameters the same as the blue curve,
but we limit the temperature to be $T\leq T_{\rm melt}$. Again, we find that we cannot match the luminosity
observed at early times in each outburst.

	To summarize, if a large fraction of the surface is heated, we find that we
can match the light curves if the inner crust is heated in the 2008 outburst
(and possibly in both outbursts). The energetics of each outburst are then similar,
with the difference in outburst energy and timescale caused by a difference in the fraction of energy
deposited in the outer crust. If instead only a small part of the neutron star surface
is heated, or if the temperature in the crust is limited by the melting temperature,
the initial parts of the outburst are too luminous to be reproduced by crust cooling
models. In that case, another explanation is needed for the early time emission.
It could be, for example, magnetospheric in origin, with the neutron
star cooling taking over at late times.

	Although we do not use a detailed model for magnetospheric relaxation in this work,
a plausible relaxation scenario for the early time trend after the outbursts is 
the untwisting model \citep[][]{b09,llb16}, which was used for explaining the transient relaxation
of the magnetar XTE~J1810$-$197 \citep[][]{gh07,ah16}.
In this model, the external magnetic field is twisted
due to sudden crustal motion, and then untwists with time. Ohmic dissipation
in the magnetosphere produces radiation in the X-ray band. The relaxation trend is determined
by the evolution of the external twisted fields, and is demonstrated
in Figure~10 of \citet{b09} or in Figure~8 of \citet{llb16}. This
may explain the early time trends in Figure~\ref{fig:fig2}. The time scale in the model is
given by $t_{\mathcal{V}}=\mu/cR_*\bar{\mathcal{V}}$, where $\mu$ is the magnetic moment,
$c$ is the speed of light, $R_*$ is the radius of the star, and $\bar{\mathcal{V}}$ is the average
threshold voltage for discharge \citep[see][for more details]{b09}.
Assuming $B=2\times 10^{14}$\,G, $R=10$\,km, and $\bar{\mathcal{V}}=10^9$\,V for SGR1627,
$t_{\mathcal{V}}$ is $10^9$\,s. Then the time for the flux to drop by an order of magnitude is
$\approx$0.02$t_{\mathcal{V}}\approx$200\,days similar to what we see in SGR1627.
Further modeling would be needed to reproduce the shape of the flux decline in detail.

\section{Discussion and Conclusions}
\label{sec:sec4}
	We analyzed X-ray data from SGR1627 taken with {\it Chandra}, {\it XMM-Newton}, and
{\it NuSTAR}. With these data, we constructed long-term cooling curves of the magnetar and
studied the properties of the star and the outbursts using a crustal cooling model.
We also studied the X-ray spectrum of SGR1627 and found evidence for
a spectral turn-over at $\sim$10\,keV.

\subsection{Hard X-ray Emission}
\label{sec:sec4_1}
	In the data analysis including the {\it NuSTAR} data,
we find evidence for hard X-ray excess in SGR1627.
There are several possible causes of this if not the emission of the magnetar.
CXOU~J163547.0$-$473739 is too far off-axis to affect the results for the small
source aperture we used for SGR1627 (Figure~\ref{fig:fig1});
only 0.25\% (2 events in this case) contamination is expected into a $R=20''$
circle at a distance of $2'$.
The stray light and the ghost rays may be a larger concern
\citep[Figure~\ref{fig:fig1}; see][for more details]{mccf+17}.
However, these backgrounds are soft, having a hardness ratio (count ratio in the
10--20\,keV and the 3--10\,keV bands) $0.12\pm0.01$,
while the ratio for SGR1627 is $0.63\pm0.25$.
With the positional coincidence
between the hard excess and the soft-band detection,
it is unlikely that the hard excess is produced by the stray light.
Agreement between the {\it XMM-Newton} and the {\it NuSTAR} spectra in the common band
further suggests that the detection may not be spurious.
Moreover, in both cases, the background collected near the source position will have similar
contamination and is subtracted in the spectral analysis. We therefore conclude that the spectral
turn-over above 10\,keV in SGR1627 may be real. However, the statistical significance for
the turn-over is not very high, and further deep {\it NuSTAR} observations
are needed to confirm the hard excess.

	In some magnetars, a rising trend in the SED at
$\gapp$10\,keV has been seen \citep[][]{khdc06}.
Theoretically, the hard X-ray emission is explained with upscattering
of soft thermal photons
to higher energies $\gapp$10\,keV by outflow of $e^{+/-}$ plasma
in the lower part of the magnetosphere \citep[][]{bel13}. In this model,
the hard X-ray spectral shape strongly depends on the viewing geometry, and
the hard X-ray excess of SGR1627 suggests that its magnetic inclination
may be modest \citep[see Figure~7 of][]{bel13}; for inclinations
less than $30^\circ$, 10--20\,keV SEDs drop with energy.
Further phase-resolved spectroscopy is required for the model
to infer the inclination more precisely.

	Observationally, \citet{kb10} found a correlation between
magnetic field strength $B$ and the degree of spectral break
($\Gamma_{\rm S}-\Gamma_{\rm H}$) in the magnetar population. Following their work, 
a break of $\sim$2.5 is expected which is the same as we measure for the source if the hard excess
is from the magnetar.
Thus if the detection is real, SGR1627 can be added to the list of hard X-ray detected magnetars
and possibly to the correlation found by \citet{kb10} for future population studies.
It is interesting to note that the transient magnetar SGR1627 shows a possible
hard X-ray turn-over in the quiescent state; perhaps other faint transient magnetars
also show the turn-over in quiescence. If so, deep {\it NuSTAR} observations of these magnetars
may be useful for population studies.

\subsection{Flux Relaxation of SGR1627 Outbursts}
\label{sec:sec4_2}
	With the new observations, we added flux measurements for SGR1627 after $\gapp$2000\,days into
the outburst relaxation in order to study the late time
relaxation trends of the magnetar using a crustal cooling model.
These observations clearly show that the quiescent flux level is lower than is assumed
by \citet{aktc+12}, suggesting that the core temperature of SGR1627 may be low as originally proposed
by \citet{kewl+03}. We further show that the light curve of this magnetar exhibits another
break at $\sim$2000\,days after the 2008 outburst, similar to what was seen in the
same source after its 1998 outburst (Figure~\ref{fig:fig2}).
We attempted to explain the flux relaxation with a crustal cooling model.
While the model can explain the relaxation trends as in \citet{dcbr17}, it is
hard to match the early trends in some other cases
with crustal cooling only. This suggests that some other relaxation
mechanisms such as magnetospheric untwisting \citep{b09}
work simultaneously at early times (Section~\ref{sec:sec3_2}).

		The locations of energy deposition in outbursts of magnetars can be different:
the outside \citep{bl16}, at shallow depth \citep[][]{kyps+06}, or
the inner crust \citep[][]{llb16}. In these models, energy is generated via mechanical failure
in the crust \citep[e.g.,][]{pp11,ll12,bl14} and propagates inside, heating the star.
Using simulations \citet{bl16} and \citet{llb16} suggested that heat deposition
due to Hall-mediated avalanches and thermoplastic failures in magnetars
should happen just below the outer parts of the crust
at a depth $\gapp 100$\,m from the surface \citep[][]{bl16}.
The thermoplastic wave propagates fast and
a long distance in their simulations, hence deposits heat deeper.
If the whole stellar surface is heated by the outburst,
our study of crustal cooling for the 2008 outburst of SGR1627 supports the scenario for the energy
deposition in the inner crust as proposed by \citet{bl16} and suggests that outburst energy
may be deposited at different depths between outbursts. However, models with energy deposited
in only a small fraction of the star or with the temperature limited to the melting temperature
do not require the energy to be in the inner crust.

	With the current data sets and models, it is hard to tell whether the whole surface,
or only a small fraction is heated and thus another mechanism (e.g., magnetic untwisting)
may operate at early times. These can be further studied by measuring and modeling spectral evolution
after outbursts. However, our crust model does not account for the spectral evolution
yet because of complex interplay between the surface thermal emission
and the magnetospheric plasma. In particular, the 1\,keV effective temperature derived
from our blackbody fits is much larger than the expected surface temperature in these models.
Furthermore, the current data quality does not allow us to
measure spectral evolution of the source during the relaxation.
Further theoretical studies and high-quality observations are needed.

	For the crustal cooling models, more outburst samples
and further modeling works will help to break degeneracy in the models between the energy deposition
profile and internal properties of neutron stars. Furthermore, measuring relaxation trends until
very late times is an important probe for studying the deep interior of neutron stars which may contain
interesting structure such as nuclear ``pasta''. More detailed studies with a larger number
of outbursts may give us new insights into the location of crustal failures,
relaxation mechanisms, and the internal matter of neutron stars.

\bigskip

This research was supported by Basic Science Research Program through
the National Research Foundation of Korea (NRF)
funded by the Ministry of Science, ICT \& Future Planning (NRF-2017R1C1B2004566).
VMK is supported by an NSERC Discovery Grant and Herzberg Award,
the Canadian Institute for Advanced Research, an FRQNT Centre Grant,
a Canada Research Chair, and the Lorne Trottier Chair in Astrophysics \& Cosmology.
AC is supported by an NSERC Discovery Grant. AC thanks the School of Mathematics, Statistics,
and Physics at Newcastle University for hospitality during completion of this work.

\bibliographystyle{apj}
\bibliography{MAGNETAR,GBINARY,BLLacs,PSRBINARY,PWN,STATISTICS,ABSORB,INSTRUMENT,COOLINGMODEL}

\end{document}